\documentclass[twocolumn,superscriptaddress,showpacs,amsmath,amssymb,pra,longbibliography]{revtex4-1}

\usepackage{graphicx} 
\usepackage{dcolumn}  
\usepackage{dtklogos}
\usepackage{bm}       
\usepackage[usenames,dvipsnames]{color}
\definecolor{URLCOL}{rgb}{0,0.52,0.83} 
\definecolor{LINKCOL}{rgb}{0.05,0.5,0} 
\definecolor{orange}{rgb}{0.6,0.3,0} 
\definecolor{CITECOL}{rgb}{0.25,0,0.48} 
\usepackage{epstopdf}

\usepackage[bookmarks,bookmarksopen,bookmarksnumbered,colorlinks,linkcolor=LINKCOL,linktocpage,citecolor=CITECOL,urlcolor=URLCOL,pdfpagemode=UseOutlines]{hyperref}

\usepackage{booktabs}
\usepackage{comment}
\usepackage{natbib}

\definecolor{TITLECOL}{rgb}{0.1,0.2,0.7} 
\definecolor{SECOL}{rgb}{0.1,0.2,0.7} 
\definecolor{CONTENTSCOL}{rgb}{0.1,0.2,0.7} 
\definecolor{SSECOL}{rgb}{0.25,0,0.48} 
\definecolor{SSSECOL}{rgb}{0.2,0.08,0.53} 
\definecolor{FINCOL}{rgb}{0.01,0.3,0.07} 

\definecolor{URLCOL}{rgb}{0,0.17,0.43} 
\definecolor{LINKCOL}{rgb}{0.05,0.4,0} 
\definecolor{CITECOL}{rgb}{0.35,0,0.48} 

\def\sss{\scriptscriptstyle\rm}

\def\bea{\begin{eqnarray}}
\def\eea{\end{eqnarray}}
\def\ben{\begin{equation}}

\def\een{\end{equation}}
\def\benu{\begin{enumerate}}
\def\enu{\end{enumerate}}
\def\bei{\begin{itemize}}
\def\eei{\end{itemize}}

\def\br{{\bf r}}

\def\n{n}

\def\sss{\scriptscriptstyle\rm}

\def\x{_{\sss X}}
\def\c{_{\sss C}}
\def\s{_{\sss S}}
\def\xc{_{\sss XC}}

\def\H{_{\sss H}}

\def\ee{_{\rm ee}}

\def\bei{\begin{itemize}}
\def\eei{\end{itemize}}
\def\beit{\begin{itemize}}
\def\eit{\end{itemize}}
\def\benu{\begin{enumerate}}
\def\enu{\end{enumerate}}

\def\br{{\bf r}}

\def\x{_{\sss X}}
\def\c{_{\sss C}}
\def\s{_{\sss S}}

\def\t{^{\tau}}
\def\xc{_{\sss XC}}

\def\H{_{\sss H}}
\def\ee{_{\rm ee}}

\def\tp{^{\tau'}}
\def\inft{\tau''}

\def\FF{A}

\def\st{_{\sqrt{\tau'/\tau}}}

\begin{document}
\title{Connection formulas for thermal density functional theory}
\author{A. Pribram-Jones}
\affiliation{Department of Chemistry, University of California, Irvine, CA 92697}
\author{K. Burke}
\affiliation{Department of Chemistry, University of California, Irvine, CA 92697}
\affiliation{Department of Physics and Astronomy, University of California, Irvine, CA 92697}
\date{\today}
\begin{abstract}
The adiabatic connection formula of ground-state density functional
theory relates the correlation energy to a coupling-constant
integral over a purely potential contribution, and is widely used
to understand and improve approximations.  The corresponding
formula for thermal density functional theory is cast as an integral
over temperatures instead, ranging upward
from the system's physical temperature.
We also show how to relate different different correlation components
to each other other, either
in terms
of temperature- or coupling-constant integrations.
We illustrate our results on
the uniform electron gas.
\end{abstract}
\pacs{31.10.+z,31.15.E-}
\maketitle

The adiabatic connection formula\cite{HJ74,LP75,GL76}, often derived via
the Hellmann-Feynman theorem\cite{H33,F39}, has been used
extensively as an interpretive and development tool in
density functional theory (DFT)\cite{HK64,KS65}.  By scaling the
electron-electron interaction with a coupling constant, 
while keeping the density fixed, one constructs a
path from the non-interacting Kohn-Sham system to
the fully interacting system of interest. This yields the exchange-correlation (XC) energy as an integral over
only a purely potential contribution.

For equilibrium systems at finite (i.e., non-zero) temperatures, 
Mermin\cite{M65} generalized the HK theorems
of ground-state DFT. Applying the theorem to the Kohn-Sham scheme of
fictitious non-interacting electrons with the same  equilibrium density,
one finds a set of thermal KS equations\cite{KS65}, in which the KS orbitals
are thermally occupied via a Fermi function\cite{PPGB14}.  The relation to the
physical system is given by the thermal XC free energy, which now
includes an entropic contribution.    The dependence of the 
Hartree and exchange energies on the coupling constant is simple\cite{PPFS11},
but the thermal correlation free energy is more complicated.
Relating scaling to the coupling constant, the thermal adiabatic connection
formula was derived in Ref. \cite{PPFS11}.

Here, we show that the adiabatic connection formula at finite temperature
can be recast as an integral over temperatures, {\em without changing the
coupling constant}.  This thermal connection formula for the XC free
energy at temperature $\tau$ is
\ben
\FF\xc\t[\n]=\frac{\tau}{2}\lim_{\inft\to\infty}\int_\tau^{\inft}\frac{d\tau'}{\tau'^2}\, U\xc^{\tau'}[n\st]
\label{Axcth}
\een
where $U\xc\t[\n]$ is the purely potential contribution to the XC
free energy, and 
\ben
\n_\gamma(\br)=\gamma^3\, \n(\gamma\br)
\een
is the usual coordinate scaling of the density introduced by Levy and
Perdew\cite{LP85} for the ground-state problem.  Thus Eq. (\ref{Axcth})
extracts the XC free energy, including both kinetic and entropic contributions,
from the interaction contribution alone.  Intriguingly, it is expressed
as an integral over all temperatures {\em higher} than the temperature of
interest.

However, just as in ground-state DFT, knowledge of {\em any} component
of the correlation energy as a functional of the density is sufficient
to determine {\em any} other\cite{FTB00}.   Thus approximations to the correlation
energy can be made for any one  of these components, and converted into, e.g.,
an approximation to the correlation free energy.  An example is the
`upside-down' adiabatic connection in which, for strongly correlated
systems, it can be advantageous to derive approximations to the kinetic
correlation energy\cite{SGS07,GSV09,LB09,MG12}.  Such formulas were written down
and collected explicitly in Ref. \cite{FTB00}, and even used to construct
accurate and approximate adiabatic connection curves.

This paper reports both the thermal connection formula and the derivation and many results for these formulas
at non-zero temperatures, with examples from the uniform gas.  Atomic units
are used throughout, with energies in Hartrees and distances in Bohr radii.

To begin, we review
only those thermal DFT concepts needed
to proceed, beginning with the Mermin-Kohn-Sham
(MKS) equations.  For a full introduction to thermal DFT, please see Ref. \cite{PPGB14}, and see Ref. \cite{E10} for an alternative perspective presented via Legendre transforms. The MKS equations closely resemble those at zero temperature, though they are complicated by temperature-dependent eigenvalues and chemical potential\cite{KS65}:
\ben
\label{FTKS1}
\left[-\frac{1}{2}
\nabla^2+v\s\t({\bf r})\right] \phi\t_{i}({\bf r})
= \epsilon\t_{i} \phi\t_{i}({\bf r})
\een
where $v\s\t(\br)$ is {\em defined} by requiring that the resulting
thermal density
\ben
n\t(\br)=\sum_i f\t_i |\phi_{i}({\bf r})|^2,
\een
matches that of the physical problem, where 
\ben
f\t_i=\left(1+e^{\left(\epsilon\t_i-\mu\right)/\tau}\right)^{-1}
\een
are Fermi occupation factors at temperature $\tau$.
The chemical potential $\mu$ is chosen to yield the desired
average number of electrons, $N$.
In the usual way\cite{FW71}, the free energy of the physical system is
\ben
A = T + V\ee + V - \tau S
\een
where $T$ is the kinetic energy, $V\ee$ the inter-electron repulsion,
$V$ the one-body potential, and $S$ the entropy.  In terms of the
corresponding KS quantities:
\ben
A = T\s + U + V - \tau S\s + A\xc,
\een
where the subscript $s$ denotes evaluation for the KS system, 
$U$ is the Hartree energy, and $A\xc$ is the XC free energy, defined
by this relation.  
All quantities can be considered as density functionals, in which
the chemical potential has already been eliminated in terms of the mean
particle number, since that is determined by the density.   
Most are also explicitly temperature-dependent.
Others, such as the Hartree and one-body energies,  are not, but their values for
a fixed potential vary with temperature via the temperature-dependent density.  
Because the density minimizes the free energy, one
finds
\ben
\label{FTKS2}
v\s\t[\n](\br) = v(\br)+ v\H[\n\t](\br) + v\xc\t[\n\t](\br)
\een
where $v\H[\n](\br)$ is the traditionally defined Hartree potential\cite{N02,PYTB14} and
\ben
v\xc\t[\n] (\br) = \delta A\xc\t/\delta \n(\br).
\een

\def\tl{^{\tau,\lambda}}
Unlike the ground state XC energy, $A\xc$ includes entropic contributions.
Here our focus is on the correlation effects, so we subtract off the
exchange contribution (which can be isolated by scaling to the
high-density (weakly-coupled) limit\cite{PPFS11}).  Then the kinetic
correlation energy is
\ben
T\c\t[\n]=T\t[\n]-T\s\t[\n],
\een
while the potential correlation energy is
\ben
U\c\t[\n]=V\ee\t[\n] - U[\n] -E\x\t[\n].
\label{Ucdef}
\een
Both these are exact analogs of  their ground-state counterparts.
But we also have correlation entropy:
\ben
S\c\t[\n]=S\t[\n]-S\s\t[\n].
\een
We write $A\c\t$ as a sum of two contributions:
\ben
A\c\t[\n] = K\c\t[\n] + U\c\t[\n],
\label{AfromKU}
\een
where 
the kentropic component is
\ben
K\c\t[\n] = T\c\t[\n]-\tau\, S\c\t[\n]
\label{Kcdef}
\een
and this combination plays a role mimicking that of the kinetic correlation
alone in the ground-state case\cite{PPFS11}.

Ref. \cite{PPFS11} introduced two important results.  The first is the
relation between coupling constant and scaling at finite temperature.
Introduce a coupling constant $\lambda$ in front of $V\ee$, which
is a positive number and consider varying $\lambda$ keeping $\n(\br)$ fixed.
The physical system has $\lambda=1$, while $\lambda=0$ reduces to the KS
system.  In a method similar to that used in the ground state\cite{B07}, 
combining Eq. (31) of Ref. \cite{PPFS11}
 with finite-temperature density scaling and the
relationship between coordinate- and interaction-scaled statistical operators yields
\ben
A\xc\tl[\n] = \lambda^2\, A\xc^{\tau/\lambda^2} [\n_{1/\lambda}],
\label{Axctl}
\een
where $A\xc\tl$ is the value at coupling constant $\lambda$, and 
on the right the density has been coordinate-scaled. This relates changes
in the coupling constant to coordinate scaling of the density,
just as in the ground-state theory\cite{LP85}.  
All components of the energy, such as the exchange,
kentropic, and potential contributions, scale in the
fashion of Eq. (\ref{Axctl}).  Because exchange is evaluated
on the KS thermal density matrix of non-interacting electrons, it
scales simply with $\lambda$\cite{PPFS11}:
\ben
A\x\tl[\n] = \lambda\, A\x^{\tau/\lambda^2}[\n].
\label{Axtl}
\een

A second important result of Ref. \cite{PPFS11} is the
conventional adiabatic connection formula.  Write this in terms of correlation alone by using the above lambda-scaling of exchange,
\ben
A\c\t[\n]=\int_0^1 \frac{d\lambda}{\lambda}\, U\c\tl[\n].
\label{Actac}
\een
This extracts the full C free energy from its potential contribution alone,
but at the price of having to integrate over the coupling constant.
This is a generalization of the formula that has proven so useful at
zero temperature\cite{LP75}.

But one can go further than this, and convert all coordinate scaling
into temperature scaling, yielding very different formulas.  
Begin with the exchange-correlation version of Eq. (\ref{Actac}) and  insert Eq. (\ref{Axctl}).
Define $\tau'=\tau/\lambda^2$, and change variables to find Eq. (1).
This is one of the central results of this paper:  The XC free
energy can be extracted from the potential-only contribution, 
as a temperature integral, not a coupling-constant integral.
This integral runs from the given temperature {\em upwards},
and so does not include information from the ground-state functional,
but rather from the high-temperature limit.

We can also generalize the adiabatic connection formula for
$A\c\t$ to arbitrary
coupling constant.  This follows precisely the derivation in ground-state
DFT\cite{LP85}.  Apply Eq. (\ref{Axctl}) to
Eq. (\ref{Actac}), insert the adiabatic connection, and identify the
potential-only piece inside the integral to find:
\ben
A\c\tl[\n] = \int_0^\lambda \frac{d\lambda'}{\lambda'}\, U\c^{\tau,\lambda'}[n].
\label{Actl}
\een
We can then generalize the thermal connection to arbitrary coupling constants:
\ben
\FF\c\tl[\n]=\frac{\tau}{2}\lim_{\inft\to\infty}\int_{\tau/\lambda^2}^{\inft}
\frac{d\tau'}{\tau'^2}\, U\c^{\tau'}[\n_{\sqrt{\tau'/\tau}/\lambda}].
\label{Acthl}
\een
This shows that we can trivially generate the coupling constant
dependence of $A\xc$ by changing the limits of the thermal integration
and scaling the density argument.  
This completes our formulas for extracting
$A\c\t$ from $U\c\t$.  These are useful when an expression (exact or
approximate) is derived for $U\c\t$, to get an expression for $A\c\t$.
Our thermal connection formula negates the need for a coupling-constant
dependence when $\tau$ is finite.

However, it can also happen that, e.g., by calculation, $A\c\t$ is known,
but it is desired to extract $U\c\t$, i.e., the reverse process.  This is
used in ground-state DFT when plotting the integrand in the adiabatic
connection formula\cite{H98,CS99,JS98}.  It is now straightforward to find this relation, by
differentiating Eq. (\ref{Acthl}) with respect to $\lambda$, yielding
\ben
U\c\tl[\n] = \lambda\, \frac{dA\c\tl[\n]}{d\lambda}.
\label{Uctl}
\een
In the special case where $\lambda=1$, we find the compact result
\ben
U\c\t[\n] = \frac{dA\c\tl[\n]}{d\lambda}\Big|_{\lambda=1},
\label{UctfromA}
\een
which is exactly analogous to the ground-state formula.
This cannot be simply rewritten without the coupling-constant dependence,
as derivatives with respect to scaling yield terms that depend on the
potential.

Since none of the components of the correlation free energy are independent,
we can also write the free energy in terms of the kentropic contribution
alone.  This is sometimes used in ground-state DFT (where the kentropy is
just the kinetic energy) to create approximations starting from the 
strictly-correlated limit\cite{GSV09}.  In our case, we begin with Eq. (\ref{AfromKU}),
inserted into Eq. (\ref{Uctl}) to yield
\ben
K\c\tl[\n] = A\c\tl[\n] 
- \lambda\, \frac{dA\c\tl[\n]}{d\lambda},
\label{Kctl}
\een
showing how to extract $K\c\tl$ at any coupling strength from $A\c\tl$.
More specifically,
\ben
K\c\t[\n] = A\c\t[\n]
-  \frac{dA\c\tl[\n]}{d\lambda}\Big|_{\lambda=1},
\label{KctfromA}
\een
which means any approximation for $A\c\t[\n]$ uniquely determines
an approximation for $K\c\t[\n]$.
But we can also regard Eq. (\ref{Kctl}) as a differential equation in 
$\lambda$, and solve for $A\c\tl$, to find:
\ben
A\c\tl[\n] = - \lambda\, \int_0^\lambda \frac{d\lambda'}{\lambda'^2}\,
K\c^{\tau,\lambda'}[\n],
\label{ActlfromK}
\een
which is the generalization of the ground-state adiabatic connection
formula in terms of $K\c\t$ to finite temperature
(Eq. 18 of Ref. \cite{FTB00}).  For the physical system:
\ben
A\c\t[\n] = - \int_0^1 \frac{d\lambda}{\lambda^2}\,
K\c\tl[\n].
\label{ActfromK}
\een
Finally, we can convert this into a thermal connection formula using the
universal rule for scaling, and changing variables, we find:
\ben
A\c\t[\n] = - \frac{\sqrt{\tau}}{2} \lim_{\inft\to\infty}\int_\tau^{\inft} \frac{d\tau'}{\tau'^{3/2}}
\, K\c^{\tau'} [\n\st].
\label{ActfromKth}
\een
This is the thermal connection formula in terms of the kentropic correlation
energy.  

It is straightforward to combine these various results to 
form relations between $K\c\t$ and $U\c\t$.  We find:
\ben
K\c\t[\n] = \int_0^1 \frac{d\lambda}{\lambda}\, U\c\tl[\n] - U\c\t[\n],
\label{KfromU}
\een
while the reverse relation is
\ben
U\c\t[\n] = - \int_0^1 \frac{d\lambda}{\lambda^2}\, K\c\tl[\n] - K\c\t[\n].
\een
We can turn these into thermal connection formulas.  For Eq. (\ref{KfromU}),
we simply use $A\c\t=K\c\t+U\c\t$ in Eq. (\ref{Axcth}) to find
\ben
K\c\t[\n]=\frac{\tau}{2}\lim_{\inft\to\infty}\int_\tau^{\inft}
\frac{d\tau'}{\tau'^2}\, (U\c^{\tau'}[\n\st]- 2 U\c\t[\n])
\een

Lastly, we discuss how these results differ from many long-known in the
plasma physics community.  In the standard approach to statistical mechanics,
the potential is given and so held fixed, and energies and potentials are
found for, e.g., fixed temperature dependence and chemical potential.
The latter can be eliminated in favor of fixed mean particle number.
But as the temperature varies, the one-body density also varies, and 
so such relations do not directly yield constraints on the exact density
functionals that are independent of the specific system under study.

Our derivation yield direct relations among density functionals.  
Because the density fixes the average particle number, this is no
longer a free variable, and so the only other dependence is on the
temperature.  Hence all derivatives with respect to temperature are
total, not partial.  By our methods, we are deriving relations that
can, for example, be used to test or construct any thermal density
functional correlation approximation, without reference to the given
system.  In fact, for inhomogeneous systems, both the chemical potential
and the one-body external potential vary in complex ways as the temperature
changes with fixed density.  The beauty of this methodology, created by
Levy and Perdew\cite{LP85} for the ground-state problem, is that those
dependencies need never be discussed.

In the special case of a homogeneous system, i.e., the uniform electron
gas or the one-component plasma, none of these effects are relevant, since
the potential and densities are constant at all temperatures.  In this
case, our formulas trivially match the long-known results\cite{DT81,IIT87,PD00},
and so the uniform case
serves as a useful consistency check.  For example, in the uniform case, our
formulas are trivially related to the standard coupling-constant
integration\cite{BH80,I82}.  
Several of the relations for the uniform gas were used
recently\cite{KSDT14} to reparameterize 
quantum Monte Carlo results for the thermal uniform gas.  
On the other hand, since the coupling
constant is defined in terms of parameters in some way averaged over
the entire system, such a treatment differs utterly from ours for
any inhomgeneous system.  For example, in our high-temperature limit,
the density retains all the inhomogeneity of the original system,
and never becomes more uniform, by definition.

\begin{figure}[htbp]
\includegraphics[width=0.9\columnwidth]{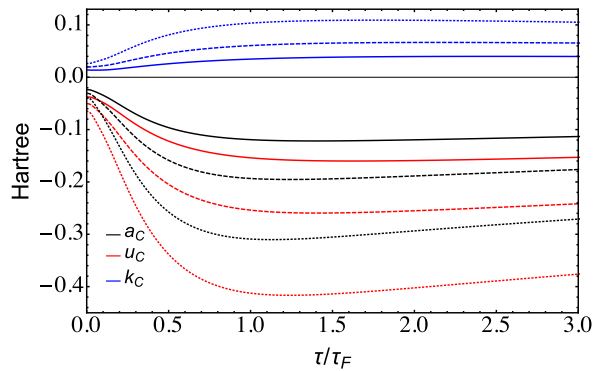}
\caption{The correlation free energy and correlation components per particle for the uniform electron gas with $r\s=2$ (solid), $r\s=1$ (dashed), and $r\s=0.5$ (dotted), as a function of the temperature in units of the Fermi temperature.  The parameterization in Ref. \cite{KSDT14} is used as a starting point, from which the correlation free energy per particle (black), the potential correlation (red), and the kentropic correlation (blue) are extracted.}
\label{acunif}
\end{figure}

\begin{figure}[htbp]
\includegraphics[width=0.9\columnwidth]{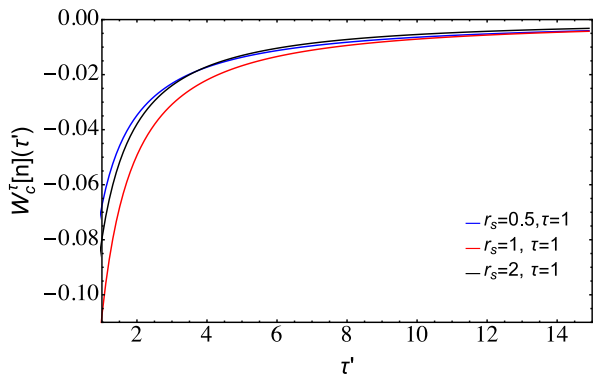}
\caption{The integrand of the thermal
connection formula for the uniform gas at various $r\s$ values with $\tau=1$.}
\label{wcunif}
\end{figure}

In Fig. \ref{acunif}, we plot the correlation free energy 
for the uniform gas, based on the recent parametrization of Ref. \cite{KSDT14}, for 
several values of $r\s$.  In every case, the free energy initially
increases in magnitude with temperature, and then slowly shrinks for
temperatures beyond the Fermi temperature.  We write the connection
formula
\ben
A\c\t[n]=\lim_{\inft\to\infty}\int_\tau^{\inft} d\tau' W\c\t[\n](\tau')
\label{Acw}
\een
where
\ben
W\c\t[\n](\tau') = \frac{\tau}{2\tau'^2}\, U\c\tp[\n\st].
\een
For the uniform gas, we divide by the volume and plot energy densities.
In Fig. \ref{wcunif}, we plot the integrand for $\tau=1$ for various
values of $r\s$.   Clearly, the curves themselves do not yield much
insight directly into the thermal correlation free energy (at least,
plotted in this way).  But we have carefully checked that
integrating
the curves from $\tau'=\tau$ to infinity
yields the $a\c\t=A\c\t/V$ values plotted in
Fig. \ref{acunif}. The smaller $\tau$ is, the more more carefully the integrand must
be approximated, especially since there is a delicate balance between the
divergence in the integrand and the prefactor of $\tau$.
But obviously approximations that expand $W\c\t(\tau')$ about
the high-temperature, high-density limit yield approximations
to $A\c\t[\n]$ via Eq. (\ref{Acw}).



To conclude, we have found an entirely new way to represent the 
XC free energy of thermal DFT, which we call the thermal connection
formula.  Unlike the ground-state adiabatic connection formula, it
relates the XC free energy to the potential contribution at higher 
temperatures.  We have also derived many other relations among the
different correlation components.  We have used these to plot the
various contributions to the uniform gas.  We anticipate these relations
playing a key role in the development of thermal density functional
approximations beyond the local approximation.

KB thanks CHE-1464795 NSF, and APJ thanks DE-FG02-97ER25308 and the University of California President's Postdoctoral Fellowship. Part of this work was performed under the auspices of the U.S. Department of Energy by Lawrence Livermore National Laboratory under Contract DE-AC52-07NA27344.

\label{page:end}

\begin{thebibliography}{10}

\bibitem{HJ74}
J.~Harris and R.O. Jones.
\newblock The surface energy of a bounded electron gas.
\newblock {\em J. Phys. F}, 4:1170, 1974.

\bibitem{LP75}
D.C. Langreth and J.P. Perdew.
\newblock The exchange-correlation energy of a metallic surface.
\newblock {\em Solid State Commun.}, 17:1425, 1975.

\bibitem{GL76}
O.~Gunnarsson and B.I. Lundqvist.
\newblock Exchange and correlation in atoms, molecules, and solids by the
  spin-density-functional formalism.
\newblock {\em Phys. Rev. B}, 13:4274, 1976.

\bibitem{H33}
H.~Hellmann.
\newblock {\em Z. Phys.}, 85:180, 1933.

\bibitem{F39}
R.~P. Feynman.
\newblock Forces in molecules.
\newblock {\em Phys. Rev.}, 56(4):340--343, Aug 1939.

\bibitem{HK64}
P.~Hohenberg and W.~Kohn.
\newblock Inhomogeneous electron gas.
\newblock {\em Phys. Rev.}, 136(3B):B864--B871, Nov 1964.

\bibitem{KS65}
W.~Kohn and L.~J. Sham.
\newblock Self-consistent equations including exchange and correlation effects.
\newblock {\em Phys. Rev.}, 140(4A):A1133--A1138, Nov 1965.

\bibitem{M65}
N.~D. Mermin.
\newblock Thermal properties of the inhomogenous electron gas.
\newblock {\em Phys. Rev.}, 137:A: 1441, 1965.

\bibitem{PPGB14}
Aurora Pribram-Jones, Stefano Pittalis, E.K.U. Gross, and Kieron Burke.
\newblock Thermal density functional theory in context.
\newblock In Frank Graziani, Michael~P. Desjarlais, Ronald Redmer, and
  Samuel~B. Trickey, editors, {\em Frontiers and Challenges in Warm Dense
  Matter}, volume~96 of {\em Lecture Notes in Computational Science and
  Engineering}, pages 25--60. Springer International Publishing, 2014.

\bibitem{PPFS11}
S.~Pittalis, C.~R. Proetto, A.~Floris, A.~Sanna, C.~Bersier, K.~Burke, and
  E.~K.~U. Gross.
\newblock Exact conditions in finite-temperature density-functional theory.
\newblock {\em Phys. Rev. Lett.}, 107:163001, Oct 2011.

\bibitem{LP85}
M.~Levy and J.P. Perdew.
\newblock Hellmann-feynman, virial, and scaling requisites for the exact
  universal density functionals. shape of the correlation potential and
  diamagnetic susceptibility for atoms.
\newblock {\em Phys. Rev. A}, 32:2010, 1985.

\bibitem{FTB00}
W.~Terilla D.~Frydel and K.~Burke.
\newblock Adiabatic connection from accurate wavefunction calculations.
\newblock {\em J. Chem. Phys.}, 112:5292, 2000.

\bibitem{SGS07}
Michael Seidl, Paola Gori-Giorgi, and Andreas Savin.
\newblock Strictly correlated electrons in density-functional theory: A general
  formulation with applications to spherical densities.
\newblock {\em Phys. Rev. A}, 75:042511, Apr 2007.

\bibitem{GSV09}
Paola Gori-Giorgi, Michael Seidl, and G.~Vignale.
\newblock Density-functional theory for strongly interacting electrons.
\newblock {\em Phys. Rev. Lett.}, 103:166402, Oct 2009.

\bibitem{LB09}
Z.-F. Liu and K.~Burke.
\newblock Adiabatic connection for strictly correlated electrons.
\newblock {\em J. Chem. Phys.}, 131(12):124124, 2009.

\bibitem{MG12}
Francesc Malet and Paola Gori-Giorgi.
\newblock Strong correlation in kohn-sham density functional theory.
\newblock {\em arXiv:1207.2775}, 2012.

\bibitem{E10}
Helmut Eschrig.
\newblock ${T}>0$ ensemble-state density functional theory via legendre
  transform.
\newblock {\em Phys. Rev. B}, 82:205120, Nov 2010.

\bibitem{FW71}
A.~L. {Fetter} and J.~D. {Walecka}.
\newblock {\em {Quantum theory of many-particle systems}}.
\newblock McGraw-Hill, New York, NY, 1971.

\bibitem{N02}
{\'A}~Nagy.
\newblock Virial theorem in the density functional ensemble theory.
\newblock {\em Acta Phys. Chim. Debrecina}, 34-35:99, 2002.

\bibitem{PYTB14}
A.~Pribram-Jones, Z.-H. Yang, J.~R. Trail, K.~Burke, R.~J. Needs, and C.~A.
  Ullrich.
\newblock Excitations and benchmark ensemble density functional theory for two
  electrons.
\newblock {\em J. Chem. Phys.}, 140:18A541, 2014.

\bibitem{B07}
K.~Burke.
\newblock The {A}{B}{C} of {D}{F}{T}, 2007.
\newblock Available online.

\bibitem{H98}
M.~K. Harbola.
\newblock Differential virial theorem for the fractional electron number:
  Derivative discontinuity of the kohn-sham exchange-correlation energy.
\newblock {\em Phys. Rev. A}, 57:4253, 1998.

\bibitem{CS99}
A.~Savin F.~Colonna.
\newblock Correlation energies for some two- and four-elactron systems along
  the adiabatic connection in density functional theory.
\newblock {\em J. Chem. Phys.}, 110:2828, 1999.

\bibitem{JS98}
D.P. Joubert and G.P. Srivastava.
\newblock Coupling-constant dependence of the density functional correlation
  energy.
\newblock {\em J. Chem. Phys.}, 109:5212, 1998.

\bibitem{DT81}
M~W~C Dharma-wardana and R~Taylor.
\newblock Exchange and correlation potentials for finite temperature quantum
  calculations at intermediate degeneracies.
\newblock {\em Journal of Physics C: Solid State Physics}, 14(5):629, 1981.

\bibitem{IIT87}
Setsuo Ichimaru, Hiroshi Iyetomi, and Shigenori Tanaka.
\newblock Statistical physics of dense plasmas: Thermodynamics, transport
  coefficients and dynamic correlations.
\newblock {\em Physics Reports}, 149(2):91 -- 205, 1987.

\bibitem{PD00}
Francois Perrot and M.~W.~C. Dharma-wardana.
\newblock Spin-polarized electron liquid at arbitrary temperatures:
  Exchange-correlation energies, electron-distribution functions, and the
  static response functions.
\newblock {\em Phys. Rev. B}, 62(24):16536--16548, Dec 2000.

\bibitem{BH80}
Marc Baus and Jean-Pierre Hansen.
\newblock Statistical mechanics of simple coulomb systems.
\newblock {\em Physics Reports}, 59(1):1 -- 94, 1980.

\bibitem{I82}
S.~Ichimaru.
\newblock {\em Rev. Mod. Phys.}, 54:1017, 1982.

\bibitem{KSDT14}
Valentin~V. Karasiev, Travis Sjostrom, James Dufty, and S.~B. Trickey.
\newblock Accurate homogeneous electron gas exchange-correlation free energy
  for local spin-density calculations.
\newblock {\em Phys. Rev. Lett.}, 112:076403, Feb 2014.

\end{thebibliography}
\end{document}